\documentclass[showkeys]{revtex4}
\usepackage{epsf,graphicx}
\begin{document}
\title[Effects of Small Molecules on Phospholipid Bilayers]{Structural Effects
of Small Molecules on Phospholipid Bilayers Investigated by Molecular
Simulations}
\author{Bryan W. Lee}
\author{Roland Faller\footnote{to whom correspondence should be addressed
at rfaller@ucdavis.edu}}
\affiliation{Department of Chemical Engineering and Materials
Science, University of California--Davis, One Shields Ave, Davis,
CA 95616, USA }
\author{Amadeu K. Sum}
\affiliation{Department of Chemical Engineering, Virginia
Polytechnic Institute and State University, 142A
Randolph Hall, Blacksburg, VA 24061, USA}
\author{Ilpo Vattulainen}
\affiliation{Laboratory of Physics and Helsinki Institute of Physics,
Helsinki University of Technology, P.\,O. Box 1100, FIN--02015 HUT,
Finland}
\author{Michael Patra}
\author{Mikko Karttunen}
\affiliation{Biophysics and Statistical Mechanics Group,
Laboratory of Computational Engineering, Helsinki University
of Technology, P.\,O. Box 9203, FIN--02015 HUT, Finland}

\begin{abstract}
\noindent We summarize and compare recent Molecular Dynamics
simulations on the interactions of dipalmitoylphosphatidylcholine
(DPPC) bilayers in the liquid crystalline phase with a number of small
molecules including trehalose, a disaccharide of glucose, alcohols,
and dimethylsulfoxide (DMSO). The sugar molecules tend to stabilize
the structure of the bilayer as they bridge adjacent lipid
headgroups. They do not strongly change the structure of the bilayer.
Alcohols and DMSO destabilize the bilayer as they increase its area
per molecule in the bilayer plane and decrease the order
parameter. Alcohols have a stronger detrimental effect than DMSO. The
observables which we compare are the area per molecule in the plane of
the bilayer, the membrane thickness, and the NMR order parameter of
DPPC hydrocarbon tails. The area per molecule and the order parameter
are very well correlated whereas the bilayer thickness is not
necessarily correlated with them.
\end{abstract}

 \maketitle
\section{Introduction}
Phospholipid bilayers have been the focus of research for a long time
due to their natural occurrence in cellular and intracellular
membranes. Detailed computer simulations of phospholipid monolayers
and bilayers have achieved a high degree of sophistication over the
last
years~\cite{feller95,tieleman97,tobias97,bandyopadhyay98,husslein98,rog01,saiz02a,patra03,patra04,leontiadou04}.
The structures of lipid bilayers in water have been determined for a
variety of
phospholipids~\cite{husslein98,mashl01,saiz02a,gurtovenko04} and the
understanding of their structural features has increased
significantly. More recently computational studies of mixtures of various
phospholipids~\cite{pandit03,balali03,vries04,gurtovenko04} and of
phospholipids with
cholesterol~\cite{tu98,smondryev00,pandit04,falck04s} have also been
reported. However, the question how such lipid bilayer membranes
interact with small molecules has to date not gained very much
attention except for a few initial studies on
alcohols~\cite{feller02,bemporad04,patra04s},
sugars~\cite{sum03a,pereira04}, and dimethylsulfoxide
(DMSO)~\cite{smondryev99,sum03b}. The influence of lipophilic polymers
on lipid bilayers has also been studied~\cite{jeng03}.

If we want to understand the interactions of cells with
their environment, we first have to understand the interaction of
cell membranes -- or phospholipid bilayers as model systems --
with molecules typical in cell environments. Cell membranes are
the first part of the cell to come into contact with any nutrient,
pathogen, or other molecule in the environment. So the
understanding of membrane interaction with small molecules is of
tremendous biological importance.

In this contribution we compare the structural influences of different
small molecules on model lipid membranes. We are focusing on
molecules with high biological relevance. These include trehalose,
alcohols and dimethylsulfoxide. Sugar molecules are obviously
nutrients to living organisms and some sugars are also known to be
cryo--protectants~\cite{crowe87a,crowe88,crowe01}. In particular,
trehalose, which is a disaccharide of glucose has been found to be
very effective in this respect. It has been pointed out that trehalose
is able to form only one internal hydrogen bond in contrast to, e.g., sucrose
which forms two~\cite{conrad99}. This leaves trehalose more susceptible
to hydrogen bonding with lipids. Recently, it has been shown that
the molecular mechanism underlying this cryo--protective effect is the
hydrogen bonding pattern of the trehalose molecules to the bilayer
headgroups~\cite{sum03a}. The sugar can ``substitute'' some of the
hydrogen bonds normally provided by water and by that stabilize
the fragile bilayer arrangement. Stabilization or destabilization here
mean that the bilayer is able to withstand harsher or less harsh
conditions in presence of small molecules compared to the pure
bilayer in water.  Trehalose is experimentally known to prevent the
lipid from undergoing a phase transition under cooling, i.e., it shifts
the main phase transition temperature significantly~\cite{crowe84b}.

DMSO is known to have some cryo--protective properties as
well~\cite{freshney87}, in addition to its anti-inflammatory,
analgesic, muscle relaxant, mast cell stimulation and collagen
dissolution properties~\cite{jacob86}. It has been reported that DMSO
strongly changes the overall bilayer structure by penetrating into the
hydrocarbon layer to a much higher degree than sugar or alcohol
molecules~\cite{sum03b}.

It has been observed experimentally that alcohols have a destabilizing effect on model
membranes~\cite{ly02,ly03s}. It has also been observed that upon addition
of alcohols, the lipid bilayer becomes thinner and the area per
molecule increases. A related industrial question is the problem known
as stuck fermentation in the wine industry~\cite{bisson02,cramer02}. A
stuck fermentation means that the yeast cells do not transform all the
available sugar into alcohol but stop at an incomplete stage. It has
been proposed that the underlying mechanism is an alcohol--triggered
structural transition in the membrane which forces trans--membrane
proteins to change conformation rendering them
dysfunctional~\cite{bisson02}. To date, there is no predictive means
to detect stuck fermentations. The first approach to understand the
mechanisms leading to stuck fermentation is therefore to study the
influence of alcohols on lipid bilayers as the direct effect on the
proteins is expected to be only secondary.
\section{Simulation Details}
\begin{figure}
\includegraphics[width=7cm]{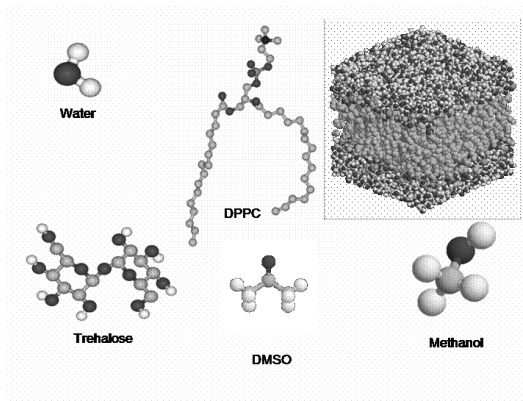}
\caption{The various molecules used in our simulation and a
typical bilayer configuration.} \label{fig:pict}
\end{figure}

In this contribution we focus on a comparison between small molecules
and their effect on the bilayer structure. The details of the
simulation models are presented
elsewhere~\cite{sum03a,sum03b,patra04s}. Let us summarize the main
characteristics. All simulations contain fully hydrated
dipalmtoylphosphatidylcholine (DPPC) bilayers with 128 molecules,
i.e., 64 per leaflet (DPPC is one of the most abundant phospholipids
in animal cell membranes). Our simulations are in atomistic detail
except for the hydrogen atoms bonded in methyl(ene) groups. These
groups are collapsed into a united atom description centered on the
respective carbon. There are at least 3655 water molecules in the
system (8958 water molecules for some systems with alcohol). All
simulations were performed at 325~K, a temperature at which DPPC is in
the biologically most relevant fluid phase. Simulations of the pure
bilayer, as well as bilayers with up to 3~wt\% methanol or ethanol,
(lipid free basis, as we do not take the lipids into account for
concentration calculations) have been performed. Note that a weight
concentration of 1.7~wt\% methanol is the same molar concentration as
2.5~wt\% ethanol. In the case of trehalose the concentrations
considered were up to 3~wt\%, simulations at higher concentrations and
different temperatures have been reported
elsewhere~\cite{sum03a,pereira04}. The DMSO systems contained
18.6~wt\% DMSO. Simulations were performed under constant temperature
and constant pressure conditions using the Berendsen weak--coupling
scheme~\cite{berendsen84}. Simulations used a time-step of
1--2~fs. Figure~\ref{fig:pict} shows all the used molecules as well as
a typical bilayer configuration.

It is worth noting that the exact simulation conditions can have a
significant effect on the structural properties. For example, changing
the cutoff for the nonbonded interactions can change the area per
molecule by as much as 10~\%~\cite{patra03}. Thus all direct
comparisons have been performed under identical conditions with a
cutoff of the Lennard--Jones interaction of 0.9~nm or 1.0~nm,
respectively.  The simulation models of the lipid alkyl tails used
either the NERD~\cite{nath98,nath00,nath01a} or the GROMOS
force-field~\cite{gromos96}; no significant differences were observed
between these two models based on simulations of the pure bilayer
system~\cite{sum03a,patra03}. This is not surprising as the two
force--fields differ only slightly in their treatment of the
hydrocarbon tails. The parameters for the head--groups are
identical. However, if we compare our areas per molecule to the data
of Pereira {\it et al.}~\cite{pereira04} who used a different model,
we see a difference of about 10\% for the pure DPPC bilayer and for
the systems with trehalose. The models for the alcohols use either the
GROMOS forcefield~\cite{gromos96} or the model suggested by
M{\"u}ller-Plathe~\cite{mplathe96b}, small differences in the density
profiles were observed. The other observables discussed here were in
all compared cases the same, a detailed comparison will be published
elsewhere~\cite{faller04sc}. Trehalose was modeled using the OPLS
model~\cite{damm97}, DMSO using the revised model by Bordat {\it et
al.}~\cite{bordat03} and water using the SPC/E
model~\cite{berendsen87}.

Electrostatic interactions have been considered using the particle
mesh Ewald technique~\cite{essmann99}. A few simulations have been
performed using a reaction field technique~\cite{allen87}. We compared
the results from the simulations using the reaction field technique
to the PME data for a representative subgroup and found no
significant differences. Simulations did not contain any salt or other
electrolytes.

We focus on three key observables to characterize the structure of
the bilayer systems in contact with small molecules. These are the
area per molecule, the membrane thickness and the alkyl tail order
parameter. The area per molecule is simply defined as the surface
area in the $xy$ plane (the membrane normal is the $z$ direction)
divided by the number of molecules per leaflet. The membrane
thickness is defined as the distance between the peaks of highest
density in a mass density profile, which roughly corresponds to the
distance between the planes of the phosphorus atoms.

The order parameter is defined in reference to NMR experiments by
\begin{eqnarray}
  - S_{CD}&=&\frac{2}{3}S_{xx} + \frac{1}{3}S_{yy},\\
  S_{\alpha\beta}&=&\langle3\cos\Theta_{\alpha}\cos\Theta_{\beta}-
  \delta_{\alpha\beta}\rangle,\qquad\alpha,\beta=x,y,z\\
  \cos\Theta_{\alpha}&=&\hat{e}_{\alpha}\hat{e}_z,
\end{eqnarray}
where $\hat{e}_z$ is a unit vector in the laboratory $z$-direction and
$\hat{e}_{\alpha}$ is a unit vector in the local coordinate system of
the tails, which involves three connected carbons C$_{i-1}$,C$_{i}$,
and C$_{i+1}$ and $\vec{e}=\vec{r}_{i+1}-\vec{r}_{i-1}$. This order
parameter characterizes the alignment of the hydrocarbon tails with the
bilayer normal. It can be used as one of the parameters (in addition
to, e.g., the area per molecule and the thickness) to characterize the
thermodynamic phase of the system. All simulations in this
contribution are in the biologically most relevant fluid phase. At
a lower temperature the order of the tails increases and this triggers a
phase transition to the gel phase.

All these observables are experimentally accessible; the order
parameter and the area per molecule can be measured by NMR. The
thickness can be measured experimentally by using electron density
profiles from X-ray scattering~\cite{nagle96}. However, especially
under the influence of small molecules it is often inferred indirectly
from the area per molecule under the assumption of volume conservation
in the bilayer~\cite{ly02}.
\section{Bilayer Structure and Influence of Small Molecules}
\begin{figure}
\includegraphics[width=7cm]{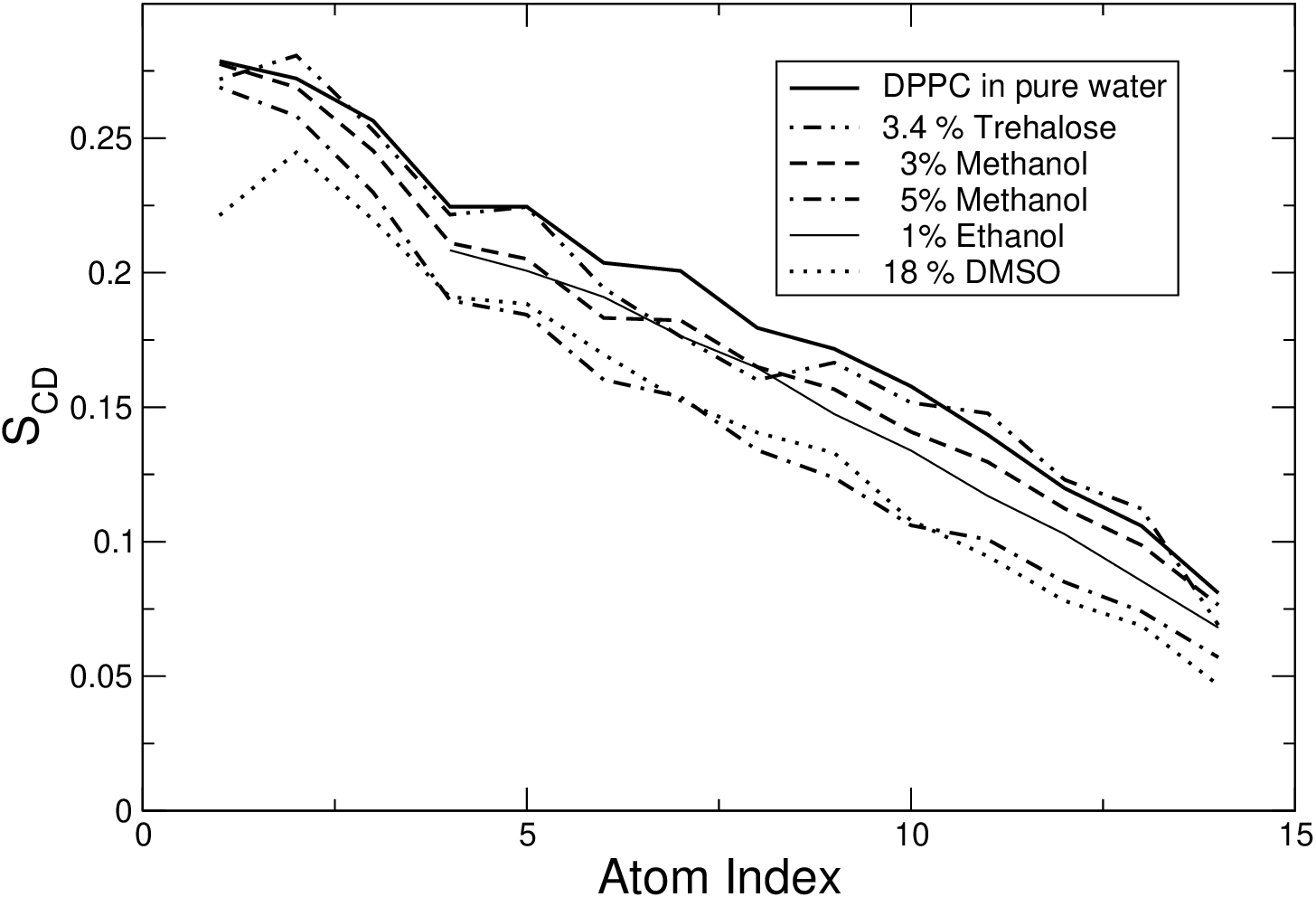}
\caption{Order parameter for DPPC at 325~K in dependence of
concentration of added small molecules. We average over both
tails. Note that the data for trehalose is different from the one
presented in~\cite{sum03a} as we are presenting data at a lower
temperature. Carbon number 1 is closest to the headgroup and
carbon number 15 is the end of the tails. } \label{fig:order}
\end{figure}

\begin{figure}

\vspace{1cm}
\includegraphics[width=7cm]{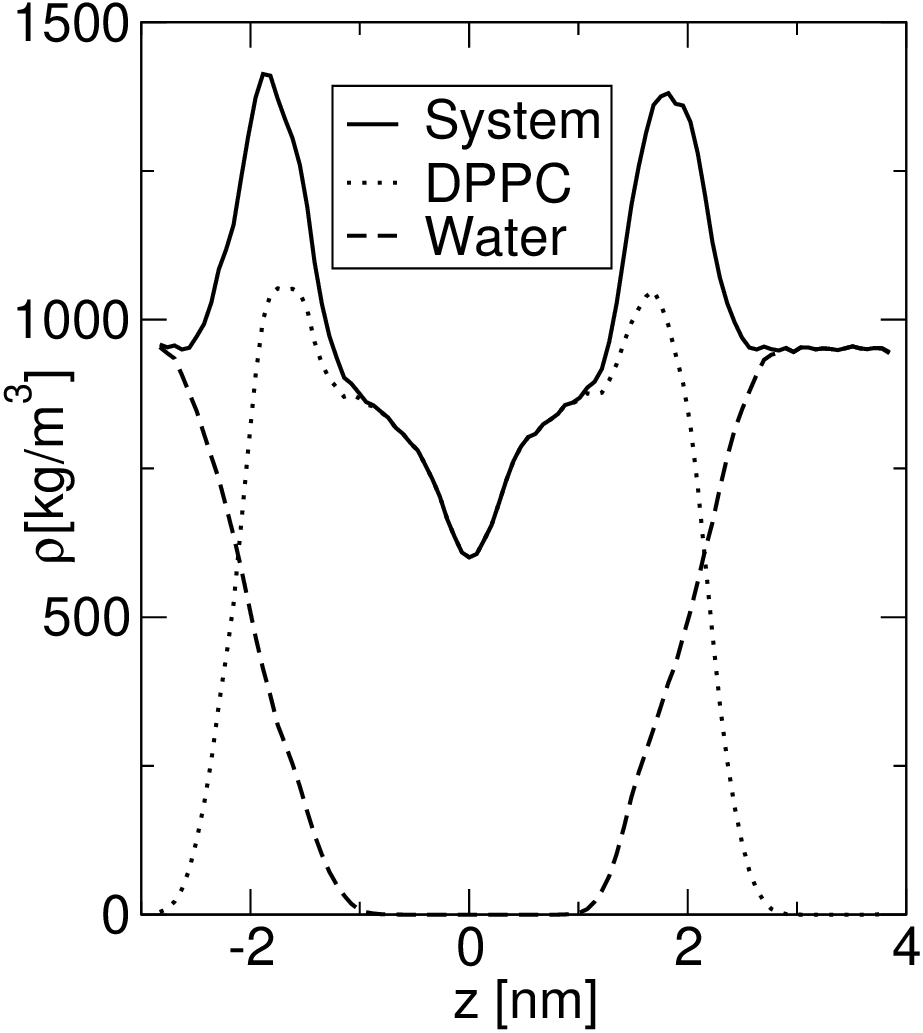}\hspace{1cm}
\includegraphics[width=7cm]{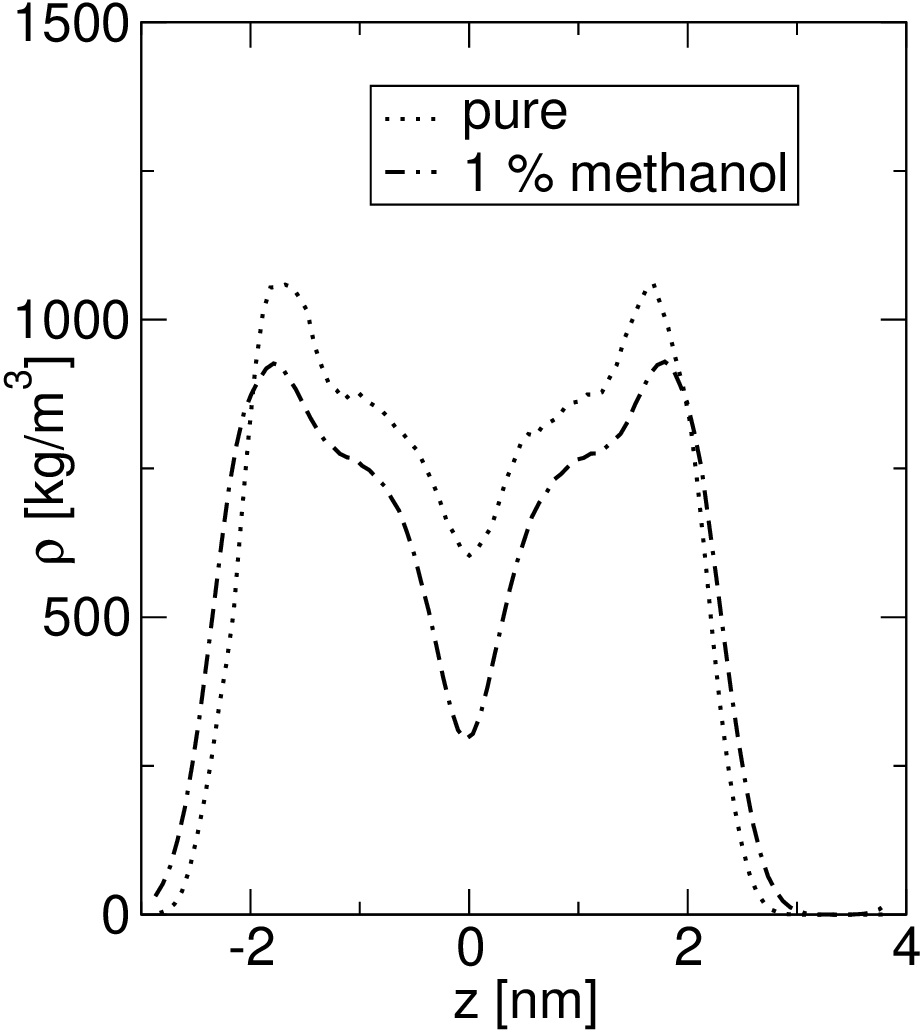}\\

\vspace{2cm}
\includegraphics[width=7cm]{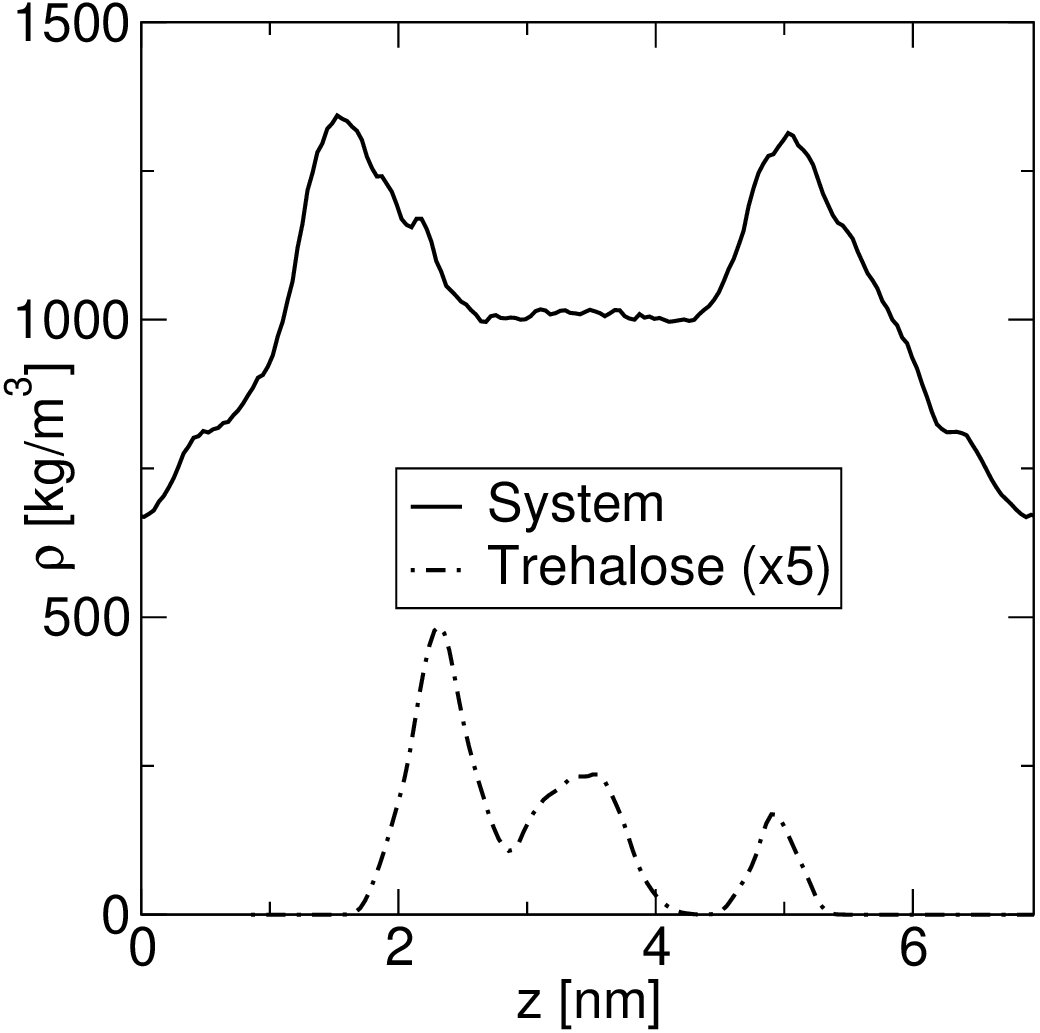}
\caption{Density profiles a) of a pure DPPC system and its
constituents b) pure system in comparison with a system containing
1~wt~\% methanol (using the model from Ref.~\cite{mplathe96b}. We
only show the density profile of the lipids here. c) system
containing 3.4~wt~\% trehalose, these curves are centered around
the middle of the water layer ($z=3.47$~nm) for clarity. Note that
the asymmetry in the trehalose distribution is a statistical
effect and not a sign of insufficient equilibration.}
\label{fig:density}
\end{figure}

At the temperature of 325~K we found the area per molecule to be
0.65~nm$^2$ for a pure DPPC bilayer which lies well within
experimentally reported values of 0.55~nm$^2$ to
0.71~nm$^2$~\cite{nagle93,nagle96,nagle00,nagle02}. The alkyl tail
order parameters for the pure DPPC system, shown in
Fig.~\ref{fig:order}, are also consistent with experimental
data~\cite{schindler75} and previous simulation
studies~\cite{tieleman97}. The area per molecule increases and the
order parameter decreases with increasing temperature in the pure
system~\cite{sum03a}. Moreover, the order parameter decreases
continuously from the head group to the end of the tails in the
bilayer center. This means that the tails in the center of the bilayer
are only very weakly aligned with the bilayer normal, especially at high
alcohol and DMSO concentrations. It is conceivable that at higher
alcohol concentrations the order may be completely lost. At very high
alcohol concentrations interdigitation between opposing
monolayers has been experimentally observed~\cite{mou94}.
Here, no interdigitation was observed as we study much lower concentrations
(cf. Fig.~\ref{fig:density}b). The thickness of pure bilayers shrinks
as the phospholipids become less ordered. This means that in the
$z$-direction -- defined to be the normal to the bilayer surface --
one finds a negative heat expansion coefficient. The overall heat
expansion coefficient is positive as the increase in area
overcompensates the loss in thickness.
Figure~\ref{fig:density}a shows the density profile of a pure DPPC
bilayer. We clearly see the self--assembly of the system into a water
region with a density very close to bulk water, an interface with a
high local density as the heavy phosphorus atoms are located in this
area, and an aliphatic region well below water density in the
center. The plane of lowest density in the middle of the bilayer where
the lipid tails from opposing leaflets meet is a symmetry plane, which
we define as the $z=0$ plane.  The interface region contains some
water and the lipid head groups, which include the choline and
phosphate groups with the density increasing up to around
1400~kg/m$^3$.  A small addition of 1.0~wt\% methanol leads to an
overall decrease of the bilayer density which suggests an increase in
the area per molecule. This is indeed observed
(cf. Table.~\ref{tab:area}), as well as an increase in the bilayer
thickness, leading to an overall decrease of bilayer density. This
result actually challenges the typical experimental assumption of
constant bilayer density often used to calculate changes in layer
thickness~\cite{ly02,ly03s}. To characterize the structure of the bilayer at least two
independent measurements out of the set thickness, area, and density
are necessary.
\begin{table}
\begin{tabular}{lrr}
Molecule & weight \% & area per molec. [nm$^2$]\\
\hline
pure water & 0 & 0.65\\
\hline
MeOH & 1 & 0.69\\
 & 1.7 & 0.69\\
 & 3 & 0.72\\
\hline
EtOH & 2.5 & 0.70\\
\hline
Trehalose & 3 & 0.64\\
\hline
DMSO & 18 & 0.72\\
\hline
\end{tabular}
\caption{Dependence of the area per molecule at 325~K under
addition of different molecules. Note that the concentrations are
weight \% lipid free basis. The Error estimates are
$\pm0.01$nm$^2$ unless otherwise noted.} \label{tab:area}
\end{table}

Experimentally it is found that low molecular weight
alcohols increase the area per molecule and decrease the order. The
experimental increase for an SOPC bilayer~\cite{ly03s} is of the order
of $\Delta A/A\approx 0.1$ at 5~wt\% methanol. Consequently, the
thickness of the membrane is inferred to be thinned out.  Measurements
using DPPC or other fully saturated phosphatidylcholines have not yet
been reported.  The only simulation study of alcohols on bilayers we
are aware of~\cite{feller02} did not allow for changes in the
box--size in the plane of the membrane as that study focused on local
structural effects and dynamics, and hence  it is not possible to directly
compare our results to those of Feller {\it et al.}~\cite{feller02}.
Here we find a linear dependence of the area
expansion on alcohol concentration as well as an increase of the
effect from methanol to ethanol which is in line with the
experimentally observed increase of the area expansion with alcohol chain
length~\cite{ly03s}. A more detailed analysis of the
alcohol--bilayer interaction is found in Ref.~\cite{patra04s}.

Bemporad {\it et al.} measured free energies of transition through a
lipid bilayer for various small
molecules~\cite{bemporad04}. Although their numbers for the
permeability are an order of magnitude too large, it is clear that
methanol faces a barrier which is almost as high as the one for water. It is
highly unlikely for the methanols to cross the
bilayer. This is very much in agreement with our density profiles as
we do not find any methanol molecules inside the layer. For ethanol,
however, a few molecules were found inside the layer in agreement with
the old Traube rule~\cite{traube1891} which says that interfacial
tension drops by a constant factor as one adds methylene groups.

The simulations with trehalose revealed only small
changes to the lipid bilayer. We found a slight increase in the
order parameter with increasing sugar concentration; this effect was
earlier found to be more pronounced for trehalose than for its isomer
sucrose~\cite{sum03a}.  We find that trehalose stabilizes and
preserves the membrane under cooling without affecting its structure. This is
understandable from the point of view of cryo--preservation and also
due to the fact taht sugars are an abundant nutrient of cells and membrane
integrity must not be strongly affected by their presence (see
ref.\cite{sum03a} for additional results and discussions). As seen
from the density profiles in Fig.~\ref{fig:density}c, trehalose does
not penetrate into the bilayer and accumulates at the
headgroups. The sugar molecules hydrogen bond to the lipid headgroups,
one trehalose molecule can bind to up to three lipids.

Pereira {\it et al.} performed simulations of a DPPC bilayer in water
with much higher concentrations of trehalose and found very similar
effects~\cite{pereira04}. The area per molecule is almost unaffected
even with 128 molecules of trehalose for 128 molecules of DPPC. The
same is true for the order parameter. Their simulation model is
different so that the area per molecule cannot be compared directly,
they get 0.58~nm$^2$ for the pure DPPC system at 325~K. But it
is clear that the effect of trehalose is very similar in two
independent studies with distinctly different models.

For the bilayer system containing DMSO, the area per molecule is
increased, but not as strongly as in the presence of methanol or
ethanol, and consequently we observe a decrease in the order
parameter. A concentration of about 18~wt\% DMSO is needed to get
the same increase in area per molecule as with 3~wt\% methanol.
Even if the molar concentrations are considered for comparison, we
obtain the same effect with 1.6~mol\% methanol versus 5~mol\%
DMSO.

As described earlier, we see an increase in the thickness of the
bilayer upon addition of methanol. We measured the thickness of the
bilayer as the distance between the planes of highest density of the
overall system. In the pure system (DPPC and water only) we obtained a
layer thickness of 3.68~nm. Addition of 1 wt\% methanol gave
4.00~nm, whereas 18 wt\% DMSO lead to a decrease to 3.24~nm. This
suggests that the change in the bilayer thickness can not be directly
inferred from the change in the area per molecule.  Again we see that
trehalose has the weakest influence with a thickness of 3.50~nm. The
errors on the thickness estimation are at most around 0.2~nm.

\section{Conclusions}
We have found, in agreement with experimental data, that trehalose has a
stabilizing effect on lipid bilayers. The bilayer structure is
essentially unchanged from the structure of the pure system even if
the temperature is lowered~\cite{crowe01,sum03a}. However, the presence of alcohol or DMSO
is contrastingly different: small concentrations of alcohol have a
detrimental effect on the bilayer structure, the influence of DMSO being somewhat
less pronounced. The area per molecule and bilayer thickness
increase in the presence of alcohols, leading to a decrease in the
overall density of the bilayer. The opposite effect is observed with
DMSO, as the bilayer density is relatively unchanged, and the area per
molecule increase leads to a thinning of the bilayer.

These results suggest that the area per molecule and the alkyl tail
order parameter are strongly anti--correlated, that is, an increase in
the order parameter is accompanied by a decrease in area per molecule
and vice--versa. Similar results have been found in a recent study on
chesterol phospholipid interaction~\cite{falck04s}. This is understood
as the area per molecule describes the two--dimensional packing
density of the headgroups and the tails. The closer the alkyl tails
are packed, the more they are ordered. It is generally accepted that
the phase transition to the gel phase -- with a strong increase in
order parameter -- at lower temperatures is due to tail packing. At
this moment, it is not clear if there is a correlation of the bilayer
thickness to the degree different molecules penetrate the bilayer.
However, even this cannot explain our observations as DMSO penetrates
the bilayer better than the
alcohols~\cite{sum03b,feller02,patra04s}. This question remains open
for the moment and further investigations are needed.

In the light of our findings, that there is no direct correlation
between bilayer thickness and area per molecule, we would recommend to
use scattering experiments to obtain the layer
thickness and not to rely on the assumption of constant bilayer volume.

We conclude that atomistic simulations are a
powerful means to study the interaction of small molecules with
model lipid bilayers and can give us detailed insight on the local
mechanisms of interactions. In order to understand the larger
scale effects, simulations with simpler models have to be
applied~\cite{shelley01,faller03c,marrink04,faller04sb}.

\section*{Acknowledgments}
One of the authors (R.\,F.) thanks H. V. Ly, M. Longo, F. Tablin,
and J. H. Crowe for interesting discussions on the experiments. We
are pleased to acknowledge the support by the Academy of Finland
through its Center of Excellence Program (I.\,V.), the European
Union through the Marie Curie fellowship HPMF--CT--2002--01794
(M.\,P.), and the Academy of Finland Grant Nos.~00119, 54113 (M.\,K.),
and 80246 (I.\,V.).  We would also like to thank the Finnish IT
Center for Science (CSC) and the HorseShoe (DCSC) supercluster
computing facility at the University of Southern Denmark for
computer resources as well as Advanced Microdevices Inc. for the
donation of processors (R.\,F.).

\section*{List of Symbols and Abbreviations}
\begin{tabular}{ll}
$A$ & area per molecule in the plane of the bilayer\\
$\delta_{\alpha\beta}$ & Kronegger Delta, $\delta=1$ if $\alpha=\beta$ otherwise $0$\\
$\Delta A$ & change in the area per molecule in the plane of the bilayer\\
DMSO & dimethylsulfoxide\\
DPPC & dipalmitoyl--phosphatidylcholine\\
GROMOS & Groningen  Molecular Simulation\\
NERD & Nath, Escobedo, de Pablo, revised force-field\\
NMR & nuclear magnetic resonance\\
PME & Particle Mesh Ewald\\
$S_{CD}$& NMR alkyl tail order parameter\\
$S_{xx},\,S_{yy}$ & components of the order parameter tensor\\
SOPC & stearoyl--oleoyl--phosphatidylcholine\\
$z$ & coordinate along the bilayer normal\\
$\Theta_{\alpha}$&angle between unit vector $\alpha$ and the bilayer normal\\
\end{tabular}

\section*{References}

\bibliography{standard}
\bibliographystyle{ffe}

\end{document}